\documentstyle[12pt,twoside,fleqn,espcrc1]{article}

\newcommand \beq{\begin{equation}}
\newcommand \eeq{\end{equation}}

\def\bfnabla{\mbox{\boldmath$\nabla$}}

\def\frac#1#2{{#1 \over #2}}

\def\half{\ifinner {\scriptstyle {1 \over 2}}
     \else {1 \over 2} \fi}

\def\simge{\mathrel{%
     \rlap{\raise 0.511ex \hbox{$>$}}{\lower 0.511ex \hbox{$\sim$}}}}
\def\simle{\mathrel{
     \rlap{\raise 0.511ex \hbox{$<$}}{\lower 0.511ex \hbox{$\sim$}}}}

\def\slashchar#1{\setbox0=\hbox{$#1$}
     \dimen0=\wd0
     \setbox1=\hbox{/} \dimen1=\wd1
     \ifdim\dimen0>\dimen1
        \rlap{\hbox to \dimen0{\hfil/\hfil}}
        #1
     \else
        \rlap{\hbox to \dimen1{\hfil$#1$\hfil}}
        /
     \fi}

\def\subrightarrow#1{
    \setbox0=\hbox{
      $\displaystyle\mathop{}
      \limits_{#1}$}
    \dimen0=\wd0
    \advance \dimen0 by .5em
    \mathrel{
      \mathop{\hbox to \dimen0{\rightarrowfill}}
         \limits_{#1}}}                           

\def\journal#1#2#3#4{\ {#1}{\bf #2} ({#3})\  {#4}}

\def\PRD{\journal{Phys.\ Rev.\ {\bf D}}}
\def\PRL{\journal{Phys.\ Rev.\ Lett.}}

\def\picture #1 by #2 (#3){
    \vbox to #2{
      \hrule width #1 height 0pt depth 0pt
      \vfill
      \special{picture #3} 
      }
    }

\def\scaledpicture #1 by #2 (#3 scaled #4){{
    \dimen0=#1 \dimen1=#2
    \divide\dimen0 by 1000 \multiply\dimen0 by #4
    \divide\dimen1 by 1000 \multiply\dimen1 by #4
    \picture \dimen0 by \dimen1 (#3 scaled #4)}
    }

\def\tilQ{\mbox{v\cdot q}}
\def\tilQ1{\mbox{$v\cdot q_1$}}
\def\tilQ2{\mbox{$v\cdot q_2$}}

\newcommand{\AmS}{{\protect\the\textfont2
    A\kern-.1667em\lower.5ex\hbox{M}\kern-.125emS}}

\title{Theoretical Conference Summary}

\author{Jean-Paul Blaizot\thanks{Member of CNRS}\address{Service de
Physique Th\'eorique, CEA-Saclay,  \\
          F-91191 Gif-sur-Yvette cedex}%
      \thanks{SPhT is a laboratory from  Direction des Sciences
de
la Mati\`ere du Commissariat \`a l'Energie Atomique, and Unit\'e de
Recherche Associ\'ee au CNRS (URA2306)}    }

\begin{document}
\maketitle


\section{Introduction}

For those who have been in the field since its infancy, this
Quark-Matter meeting has been one
long waited  for. And it will certainly  remain as the one marking the
beginning of a new era in the study of ultra-relativistic
heavy  ion collisions. The first results obtained at RHIC were presented, 
and the variety  and
quality of the data  confirm  that the potential of the machine
is truly fascinating. 

It is clearly too early to draw definite conclusions, or
to make firm statements about how new data fit theoretical
expectations.  To a large extent the results which have been shown at this 
conference do not change
qualitatively the
picture that has emerged from the analysis of nucleus-nucleus 
collisions performed until
now. However there are a few instances where natural extrapolations do not
seem to work. Furthermore,  intriguing new features have been revealed, 
indicating that we are perhaps seeing first glimpses of non trivial QCD
behavior in the bulk of nucleus-nucleus collisions.

Aside from the many discussions centered around
the new data and their interpretation, exciting progress in theory have been
reported, and there is no doubt that some of these will 
inspire further work. While there remains a
significant gap between what theorists can calculate from ``first
principles'' and the modeling of nucleus-nucleus collisions,
several developments indicate that it is diminishing, which is
very encouraging.

The outline of the talk is the following. I start by reviewing
progress reported on
  first principle
calculations of quark-gluon plasma properties.  Then I proceed
to a discussion of  the ``standard picture" of nucleus-nucleus collisions 
as it
has been developed in
the last years, and comment on several stages of the collisions: early
phase, expansion, freeze-out. In particular it will be interesting to see to
which extent the new RHIC
data confirm this picture; this will be also the occasion to discuss
interesting theoretical developments in the calculation of the ``initial"
 distributions of partons playing an essential role in the collisions. 
Then I turn to a discussion of specific observables. 

In order to
save space, the various talks, and the corresponding authors, will just be 
quoted in the main text and not referred to in the final bibliography. 

\section{What is the quark-gluon plasma?}

Part of the difficulty in identifying the formation of a 
quark-gluon plasma  in nuclear collisions is linked to the difficulty to 
calculate its detailed
properties from the first principles of QCD, and to 
have a good
physical picture of confinement or deconfinement. Progress in this 
direction are
constant though, and the gap between what can be calculated from QCD 
and what can be
measured, although it is still large, is diminishing.

Much of our detailed qualitative and quantitative knowledge of the
thermodynamics of QCD is coming
from lattice calculations. These calculations, reviewed here by  Karsch, have
steadily improved over the years and the accuracy of the present
results is truly
impressive. For the first time, we were presented accurate results
for the transition
temperature for the full QCD, with dynamical quarks, and the
predictions following various
treatments of the fermions
   are now close to each other
\cite{Karsch:2000kv}. The temperature depends on the number $N_f$ of 
light flavors and is
$T_c=(173\pm 8)$ MeV for $N_f=2$ and
$T_c=(154\pm 8)$ MeV for $N_f=3$ (statistical errors only);
for the physically interesting situation of one heavy flavor  and
two light ones $T_c$ is
close to its value for the $N_f=2$ case. It is also found that the
dependence of the equation of state on $N_f$ is  dominated by that of 
the ideal gas: the
pressure divided by the ideal gas pressure when plotted as a function
of $T/T_c$ appears to
be an almost universal curve. Finally, the energy density of the
quark-gluon plasma near 
$T_c$ is estimated to be
$\epsilon\sim 700$~MeV/fm$^3$ with 40\% uncertainty.

A quantitative understanding
of the slow approach to
the ideal gas limit that is seen in the lattice data is now emerging 
from continuum
analytical calculations.  Early suggestions
\cite{Peshier:1996ty}
that the quark-gluon plasma behaves as a gas of weakly
interacting
quasiparticles have been  put on   firm theoretical grounds using weak
coupling techniques  and various  resummations. The main effect of such
ressummations is a renormalization of quasiparticle 
properties, after which quasiparticles remain  weakly interacting degrees
of freedom. As  reported  by Iancu, this
picture is supported by a first principle calculation of  the entropy 
of the quark-gluon
plasma~\cite{Blaizot:1999ip}.  This analytical approach,
as  well as that of screened perturbation theory developed in
\cite{Andersen:1999fw}, is complemented by the techniques of
   dimensional reduction presented by Schr\"oder~\cite{Kajantie:2001iz}.

While massive quasiparticles may be the relevant degrees of freedom
at high temperature,
what happens near the phase transition remains unclear.  Pisarski 
suggested that
degrees of freedom  associated with the Polyakov
loop could play an important role. Let us recall that the
expectation value of the Polyakov loop:
$$l({\bf x})=(1/N){\rm Tr}\,{\rm P}\exp\left\{ig\int_0^{1/T}\, A_0({\bf
x},\tau){\rm d}\tau\right\}$$
constitutes, in the absence of dynamical quarks,  an order parameter
for the deconfinement transition: at low temperature the system
is invariant under 
$Z_3$ transformations and $\langle l\rangle=0$, while
$Z_3$ symmetry is spontaneously broken in the deconfined phase, which
is signalled by a non vanishing
   $\langle l\rangle$ \cite{Svetitsky:1982gs}.   An effective 
potential can be constructed
to control  the variation of the
order parameter. Pisarski argued that in the vicinity of the phase
transition the variation
of the pressure is dominated by that of this effective potential, and
he suggested some
observable consequences related to expected large event-by-event 
fluctuations in the
average pion momentum~\cite{Pisarski:2000eq}. An alternative picture 
of confinement,
involving a priori different degrees of freedom from those just 
discussed, is provided by
the picture of a dual superconductor. In this picture, confinement is 
associated to  a
monopole condensate.  Di Giacomo presented latest results of lattice 
calculations which
provide evidence for this mechanism using various abelian projections.

The studies of the heavy quark free energy, most commonly referred to 
as the heavy
quark potential, are
of direct relevance for heavy ion physics since they condition our 
understanding
of heavy quark bound states in a quark gluon plasma. The heavy quark 
free energy can be
obtained from an analysis of the correlator of Polyakov loops:
$\langle L(\vec x)L^\dagger(\vec y)\rangle\propto \exp\{-V(r,T)/T\}$ with
$r=|\vec{x}-\vec{y}|$. The potential has been  studied below and above $T_c$
and as a function of
the quark mass. Below $T_c$, and for infinite quark mass, the 
potential is confining. When
dynamical quarks are included, string breaking occurs and beyond a 
certain distance the
force between heavy quark vanishes; in other words,  the potential 
goes to a constant at
large
$r$. Systematic studies were presented by Karsch and further 
discussed by Petrecky.
It was reported
that close to $T_c$, string breaking and screening occurs already for 
distances as
small as $r\simeq 0.3$fm.

Some progress has been reported in two domains where it has been slow 
because the
problems are notoriously difficult. One concerns the relation between
Euclidean or imaginary
time quantities that one evaluates on the lattice and the real time
information that one needs
to describe various physical phenomena.   Hatsuda suggested the use of 
the method of maximal entropy to this
goal \cite{Asakawa:2000tr}, and he  presented model examples
showing how the method
works. Ultimately one would like to use such methods to study how 
hadronic spectral
densities vary with temperature, but one is still far from that goal.
  The other progress concerns 
calculations at finite baryonic
chemical potential. Miyamura presented
studies of the  response of physical
observables, such as
  meson masses, to small variations of the chemical potential $\mu$ around
$\mu=0$. Such calculations are possible because they involve the evaluation of
response functions for
vanishing chemical potentials, and this can be done on the lattice.

Latest developments in the study of the phase structure of cold dense matter
where reported
by Rajagopal who argued that if quark matter exists in the core of 
neutron stars
  then this matter should be  a color superconductor.
He discussed the
possible existence of a crystalline phase of such matter
\cite{Alford:2001ze}, and its
possible  role in neutron star glitches.
The influence of specific features of the equation of
state on neutron star properties
   were reviewed in more generality by Prakash.

\section{NUCLEUS-NUCLEUS COLLISIONS. THE STANDARD PICTURE}

The vast body of data accumulated over the last 15 years draw a
picture of ultrarelativistic
nucleus-nucleus collisions that may now be referred to as the ``standard
picture''.  The main
qualitative changes that are commonly expected as one moves up from SPS to 
RHIC energy are the following: a smaller
crossing times of the  two nuclei (about a tenth
of a fermi at RHIC), a larger initial energy density (by about 70\% at 
present energy),
larger freeze-out volumes  and longer lifetimes  (these expectations however
do not seem to be supported by the preliminary RHIC data), a slightly
larger  freeze-out temperature
and a smaller baryonic chemical potential
$\mu_B$ (in central rapidity region); finally quark  and
gluon  degrees of freedom should
play an increasingly important role.

I shall now make a few remarks related to the various stages of the 
collision. I start
from the last stages, namely the freeze-out, and finish with the 
initial conditions.

\subsection{Freeze-out and hadronization}

The data
indicate that matter
at freeze-out can be described by equilibrium distributions: particle
ratios are well fitted with two parameters, a
temperature $T$ and a baryon chemical
potential  $\mu_B$. As Rischke reminded us, this simple feature of 
chemical freeze-out is
supported by many analyses, covering a wide range of data, from SIS 
to CERN. Redlich pointed out that
the variation of the freeze-out parameters with the beam energy is such that
the energy per
particle remains approximately constant,
$E/N\approx 1$GeV \cite{Cleymans99}.    This picture seems to work equally
well at RHIC.    An interesting
question addressed by Rischke and  Brown in their talks is how the 
in-medium modifications
of hadron properties (mass, width)
  change the  freeze-out parameters.

Some understanding on how the apparent equilibrium state is reached
is provided by detailed microscopic  calculations which show
that there may be enough final state interactions to drive indeed the system
towards equilibrium (see the talks by Bass and Bravina).  
Calculations combining hydrodynamics
with quantum molecular dynamics  exhibit clearly the collective flow of matter
at the thermal freeze-out; these calculations show that  particles with
small  cross sections  (such as
the
$\Omega$) are not much affected by this collective flow, and have
accordingly  a lower apparent temperature (in agreement with what is seen
in the data).  Whereas such evidence for  collective flow persists at
RHIC (there are indications that the radial velocity is larger at 
RHIC than at the SPS, see for instance the summary by Nu Xu), the
data on HBT radii presented at the meeting lead to a  puzzling 
situation: radii are not as
big as expected, and more surprisingly, the ratio
$R_{out}/R_{side}$ seems to be smaller than 1, contradicting the 
common interpretation of
$\sqrt{R^2_{out}-R^2_{side}}$ as the lifetime of the emitting source
(this issue is further discussed in the summary by Panitkin).

In spite of the success of some detailed microscopic calculations, 
the fact that
particles ratios can be described by equilibrium distributions remains
rather puzzling.  Since similar analysis apply equally well 
to particle production
in  $e^+-e^-$ collisions \cite{Becattini:1996if}  
the apparent equilibrium
observed at chemical freeze-out may have  nothing to do with final 
state interactions. It
could  point  to some universality of the statistical properties of 
hadronization.  But even
if it is not fully understood, the picture is so  successful that it has
been extended to the production of 
rare particles, such as the
$J/\Psi$ \cite{Gazdzicki:1999rk} to which I come back at the end.

\subsection{Models and degrees of freedom}

As emphasized by Bass, there is no single space-time model describing
the entire collision
process.  This is because different degrees
of freedom enter different stages of the reactions, each stage 
requiring therefore a
specific treatment. An exception is  hydrodynamics where changes of 
microscopic degrees of
freedom are not treated explicitly but only collectively through 
changes in the equation of
state. Models which use hadronic degrees of freedom, such as those 
based on molecular
dynamics,  are designed for the late stages of the collisions. They
can be combined with hydrodynamics  to get a  better description of 
the freeze-out (see the
talks by Bass,  Bravina, Teaney). The models for the initial stages focus  on
partonic  degrees of freedom (see the talks by Eskola, Wang). Finally,  
aside from models which
aim at providing a complete space-time picture of the collisions, 
there are  models,
such as those  based on Gribov-Regge phenomenology like  the dual 
parton model, which  set
up  calculations in momentum space (see the talk by Werner).

K.~Eskola  showed that, when suitably  extrapolated to the present 
RHIC energies,
most models tend
to overpredict the multiplicities. There are large theoretical uncertainties
in each of the models, which means that they can be adjusted. But 
further measurements will
provide more stringent constraints. This is the case for instance of 
the measurement of the
transverse energy $dE_T/d\eta$, as emphasized by Eskola and Dumitru. 
It is not the place
here to  go into a detailed
comparison between models, but I find it interesting to observe that some
discrepancies between predictions could  be traced back by 
Eskola  to
differences  in the structure
functions used as inputs. That such a detailed microscopic
information could now matter for predicting the
value of a global quantity such as the multiplicity density is 
revealing of a profound
change in the field.

It has been expected that at RHIC energies, the effect of hard
or semi-hard
interactions should become important, with a substantial fraction of
particle production involving minijets at an early stage of the 
reactions.   Much progress
has been made in using perturbative QCD to estimate such minijet 
production  (see e.g.
\cite{Eskola:2000my}) and model predictions have been  reviewed by 
Eskola and Wang.  While
the role of minijets in the bulk of the reaction dynamics is still 
unclear, it has been
argued that the data provide evidence for hard scattering. One 
argument which was
presented relies on the following formula for the total
multiplicity:
$$
\frac{{\rm d}N}{{\rm d}\eta}= a\cdot N_{part}+b\cdot N_{coll},
$$
where $N_{part}$ is the number of participants and $N_{coll}$ the 
number of binary
collisions. There is evidence at RHIC for a sizeable value of  $b$ (see for
instance the talk by Milov).  One is used to
associate the term proportional to
$N_{part}$ to soft physics, as suggested by the ``wounded nucleon'' model
\cite{Bialas:1976ed}: once a nucleon has collided it becomes ineffective
   in producing particles in a subsequent collision, so that the 
multiplicity is just
proportional to the number of
participants.  However, it should be emphasized that while
a term proportional to $N_{coll}$ emerges naturally from  hard 
scatterings, there is also a
soft contribution proportional to $N_{coll}$, whose magnitude 
increases with energy
\cite{Capella:1994yb}. Therefore more detailed studies are needed to 
properly quantify the
importance of hard scattering in global observables. I shall end this
discussion by  noting an interesting attempt to relate the growth of the
multiplicity from peripheral to central  collisions to that of the
saturation scale  characterizing the parton
distributions~\cite{Kharzeev:2000ph}.

\subsection{Initial conditions}

This brings me to discuss exciting theoretical
advances which may lead to a complete determination of
that part of the nuclear wave functions which is relevant in
collisions at high energy. As a particular outcome, one may be able to
  relate global observables such as  the
total transverse energy  and the  multiplicity to  ``initial 
conditions'' which can be
calculated from first principles.
The  developments that I am referring to exploit a simplicity which 
emerges  at high
energy and for large  nuclei. As
recalled nicely by  Son at the beginning of his talk,  these are 
conditions which bring the
system  into a weak coupling regime   where one can reliably 
calculate.  Of course, in
actual collisions, one may
not quite be in this ideal regime. But the ideal case provides nevertheless
an extremely valuable reference which is worth studying in detail.

The picture being actively discussed involves the notion
of  a ``saturation'' scale $Q_s$ \cite{Mueller:1999wm}. Leaving aside
many subtleties, let me just mention that $Q_s$ is the scale at which 
QCD non linear effects
become important. It may be determined by an argument similar to that
used in high temperature QCD to determine the scale ($\sim g^2T$) at 
which perturbation
theory breaks down (see for instance \cite{Blaizot:2001nr}). At that point
the two terms in covariant derivatives become of the same order of
magnitude. Equivalently,  we expect non linear effects to become important
when
$\partial^2\sim \alpha_s \langle A^2\rangle$. Since the relevant 
dynamics is in the
transverse plane, the magnitude of the gradient is fixed by the 
transverse momentum
$p_T$. As for the  magnitude of the gauge field fluctuations they can 
be estimated from  the
particle number density in the transverse plane, i.e. $\langle A^2\rangle\sim
{N_g(p_T)}/{R^2}$, where $N_g(p_T)$  is the total number of
gluons  with transverse momentum of order 
$p_T$ and $R$  is the radius of the 
nucleus.  One then
obtains the saturation scale $Q_s$ as the value of $p_T$ which obeys:
$$
p_T^2\propto\alpha_s\frac{N_g(p_T)}{R^2}.
$$
This scale  has been obtained in various other ways. For instance
in~\cite{Blaizot:1987nc} it was determined by looking at the
apparition of non linear effects  in the
evolution equation for the
structure functions \cite{Mueller:1986wy}.

We note from the expression above that the number
   of partons occupying a small disk of radius $1/Q_s$ in the
transverse plane  is
$\sim 1/\alpha_s$. Such a large occupation number also characterizes 
the place where
unscreened longwavelength (magnetic) fluctuations cause perturbation 
theory to break
down at high temperature.  In such conditions of large number of 
quanta, classical
field  approximations may become
relevant to describe the nuclear wavefunctions. Such an approach, 
initiated by Mc Lerran
and Venugopalan \cite{MCLRV94},  has led recently to very interesting 
developments which have been discussed in the talks by Venugopalan 
and Kovchegov.

Of course our ability to calculate relies ultimately on $Q_s$ being 
small. It is
interesting in this respect to note that the formula above indicates that
$Q_s^2\propto\alpha_s A^{1/3}$, where $A$ is the number of nucleons in
the nucleus (this follows from the fact that ($N_g(p_T)\sim A$ and $R\sim
A^{1/3}$).  Thus $Q_s$ is large for (asymptotically) large nuclei.
Furthermore, because the number of gluons increases  with energy, $Q_s$
increases with energy. However, the condition $Q_s\gg \Lambda_{QCD}$ 
is only marginally
realized at  present, various estimates giving
$Q_s^2\sim 2$~GeV$^2$.

Until now we have been discussing properties of the nuclear 
wavefunctions. It may be
argued that the partons which get freed in a nucleus-nucleus 
collisions have typical
transverse momenta of the order of $Q_s$, and originate from 
collisions of gluons which,
in the  initial wave functions,
have also transverse momenta of order $Q_s$ \cite{AHMueller99}.
Those partons get freed in a short time, of order $1/Q_s$ and can
collide with partners with
comparable rapidity. The subsequent evolution of this system of 
partons was discussed by
Son, who identified several characteristic  regimes in this 
evolution, controlled
by well defined powers of the strong coupling constant
\cite{Baier:2001sb}. Son also  pointed out the essential
role  of inelastic processes in the approach of the system towards equilibrium.

\section{SPECIFIC SIGNALS}
I come now to the discussion of specific signals, that is, of
observables which are believed
to carry information about the state of matter produced at various
stages of the collisions,
and reveal some of its characteristic properties. It is by combining
information coming
from all these different signals that one can reconstruct the history of
a collision and possibly get evidence for the production of a
quark-gluon plasma. Most of
these signals have been well studied at lower energy. But, as we
shall see, the first RHIC
results already provide new perspectives on some of these ``old
subjects'';
furthermore new signals, such as jet quenching, are emerging.

\subsection{Fluctuations}

The many particles produced in high energy collisions allow for 
event-by-event analyses,
and the theory to exploit this information is developing. There are
many sources of
``interesting'' fluctuations that have been discussed, in particular those
occurring  near a phase transition (see the talk by Gavin), or close to a
critical point, or those due to the
formation of a ``disoriented chiral condensate''.  Much studied 
recently are the
fluctuations in conserved  quantities \cite{Asakawa00},  related to 
correlators of conserved currents  (see the talks by Koch, Asakawa,
Stephanov; see  also the related talk
by Pratt on ``balance functions''). Relevant conserved quantities are
the electric charge, the baryon number or strangeness. The motivation 
for focusing  on such
fluctuations is that some of the ``primordial'' fluctuations in 
conserved quantities  may
survive hadronization, and may therefore  reveal interesting 
properties of matter produced
in  the early stages of nuclear collisions.

\subsection{Elliptic flow}

The azimuthal anisotropy of the distributions of produced particles
is now commonly
characterized by the Fourier coefficient
   $$v_2=\langle\cos 2\varphi\rangle,$$ where $\varphi$ is the
azimuthal angle measured
with respect to the collision plane. A positive $v_2$ indicates that
the particles are
emitted preferentially in the collision plane, as was predicted
almost 10 years ago~\cite{Ollitrault:1992bk}.
The effect is now
beautifully verified and is
among  the results much discussed at this conference.
The fact that the flow decreases linearly with increasing centrality
constitutes perhaps the first instance where  hydrodynamics is seen to
work   quantitatively.

The main reason why  the distributions
are not isotropic, is
that elastic collisions change the directions of the particles. If
these collisions are
sufficiently frequent, pressure builds up and a collective flow  occurs
preferentially along the direction of the largest pressure gradient.
Thus the elliptic
flow carries information about the pressure and hence on the equation
of state. Detailed
calculations show however that the dependence on the equation of
state is rather weak
(see e.g. the talk by Teaney). It has been argued that it is
necessary for the pressure to
build up at early times to have a maximal effect \cite{Kolb:2001fh}. What
is meant by ``early'' time
however is not immediately obvious. One should keep in mind that the
flow takes time to
develop, a time typically of the order of the size of the system.
Furthermore, it is easy to
see that as one decreases the time at which the hydrodynamical 
motion starts, the
pressure increases but
that does not lead necessarily to an increase of the flow; indeed  the  inertia
in the hydrodynamical  equation
increases proportionally to the pressure: $w d{\bf v}/dt=-{\bfnabla} P$ with
$w=\epsilon+P$. In fact detailed calculations show that no visible
variation of $v_2$
follows when the initial time is changed from 1 to 2 fm/c
~\cite{Ollitrault:1992bk}.

Hydrodynamics appears to explain nicely the variation of $v_2$
with $p_T$ at small $p_T$
($p_T<1$ GeV), as  well as  the mass dependence of the effect
\cite{Huovinen:2001cy}. It is interesting to observe that this effect of
the mass results
from an interplay between radial flow and elliptic flow, so that the
existence of elliptic
flow may be taken as evidence for  a strong radial flow, as emphasized
by  Huovinen.
    Of course at sufficiently large $p_T$ the effect should disappear,
because the  elastic
cross-section drop and the high $p_T$ particles stop to behave
hydrodynamically. However the
apparent saturation  seen in the data around  $p_T\sim $
2 GeV remains to be understood.

A distinct  mechanism for azimuthal anisotropy was proposed by X.-N. Wang
\cite{Wang:2001fq}. High
$p_T$ particles lose energy when traversing a quark gluon plasma because
of induced
radiation. Since the energy loss increases with the length of
traversed material, one expects
this effect to soften the
$p_T$ distribution in the direction perpendicular to the collision
plane. This mechanism  probes therefore directly the anisotropy of the
spatial distribution.  In this picture,
$v_2$ is sensitive to jet quenching, and  X.-N. Wang  presented
indeed calculations from which he argued that the amount of quenching
needed to explain the
observed magnitude of
$v_2$ at large $p_T$ is comparable to that needed to explain the
$\pi_0$ spectra (see below the discussion of jet quenching).

\subsection{Strangeness suppression (?)}

   We have evidence that strangeness is to a large degree  ``equilibrated''
in nuclear
collisions.  Because the time scales for equilibration are a priori 
much smaller in a
quark-gluon plasma than in a hadron gas, this is sometimes taken as 
evidence for quark-gluon
plasma formation. However, it was shown by Greiner, following 
arguments presented by Rapp
to explain antibaryon yields, that  multi-mesonic  reactions  could 
produce antihyperons on
very short time scales.  Another mechanism for strangeness 
equilibration (at lower energy),
involving the in-medium modifications of hadron properties,  was 
proposed by G.E Brown \cite{Brown00}.

Redlich emphasized that some form of strangeness enhancement already appears
in p-A collisions, and
argued that the connection of strangeness enhancement to the
formation of a quark-gluon
plasma remains unclear.  Statistical models
which describe so well the
freeze-out transition involve constraints on the volume over which
matter can be
equilibrated.  This, added to the need to implement  conservation 
laws exactly in small
systems, leads to a change of perspective:  what is usually regarded 
as strangeness
enhancement in nucleus-nucleus collisions should perhaps better be 
viewed as  strangeness
suppression in proton-proton collisions. Redlich showed that this 
picture accounts  well for 
  the data on multistrange hadrons obtained by WA97 \cite{Hamieh:2000tk}.
However, it does not seem to be compatible with  the abrupt  onset of
strangeness enhancement reported by the  NA57 collaboration (see the 
talk by Carrer).

\subsection{Dileptons and photons}

Our present understanding of the various experiments on dilepton and photon
production has been reviewed by  Gale. The enhancement of low mass
dileptons seems to be
well accounted for by detailed calculations
of the spectral function of the
$\rho$ meson in a dense medium. The main question of whether the
effect can be interpreted as
a manifestation of chiral symmetry restoration remains opened, and
will perhaps be answered
by the expected new precision data.

The rate of  direct photon production has been calculated in  QCD  at
2-loop order, and   Gelis discussed the difficulties of such
calculations.  When
these rates are combined with an hydrodynamical model for  the evolution of
the system, a good fit to the WA98 data is obtained for parameters
compatible with quark-gluon
plasma formation (see the talk by Srivastava). However, as discussed 
by Gale, this
interpretation is not unique.

There has been also discussion of the intermediate mass dileptons, where
recent calculations suggest  that  thermal contributions may compete
with open charm \cite{Rapp:2000zw}.
Clearly a direct measurement of open charm would help clarifying the issue.

\subsection{$J/\Psi$ enhancement (?)}

The study of $J/\Psi$ production in nucleus-nucleus collisions is one
of the highlights of
the CERN/SPS program, and beautiful data on various aspects of this 
phenomenon have been
collected over the years  by the  NA50 collaboration.
Unfortunately, as we were reminded by J. Qiu, our theoretical
understanding of $J/\Psi$
production in hadronic collisions is still limited, even in
kinematical conditions
were QCD calculations could be carried out reliably. The difficult
part concerns the
evolution from the small $c\bar c$ pair, whose production can be
calculated perturbatively,
to the fully developed
$J/\Psi$ meson. Qiu provided a picture for the transition from the
pre-$J/\Psi$ partonic
state to the physical
$J/\Psi$ in which  the multiple scattering of the $c\bar c$
pair with the nuclear medium may increase  its invariant mass above
the $D\bar D$ threshold,
leading  eventually to
$J/\Psi$ suppression. This provides a mechanism for nuclear
absorption which seems also capable to
account for some of the  anomalous suppression, but fails to
reproduce  the data at large
transverse energies.

In fact, what happens at large transverse energy is likely to be
related to the transverse
energy fluctuations \cite{Blaizot:2000ev}. As reported by  Dinh, the data
can be explained by a simple model in
which the
$J/\Psi$'s are suppressed whenever they would be produced in a region
where the local energy
density exceeds some critical value. One outcome of this model is
that it is not correct to
associate the two structures in the pattern of
$J/\Psi$ suppression to the successive melting of the charmonium
resonances, first the
$\chi$ and then the $J/\Psi$. Note that  transverse  energy
fluctuations would also lead
to increased comover absorption; however the effect is much weaker in 
this case, as
shown by Capella.  To
account for the data, one needs to assume that  all the
$J/\Psi$'s are destroyed at large
$E_T$. The analysis presented by Dinh relies on the assumption that 
what suppresses
the
$J/\Psi$ is a local
phenomenon, related to the local energy density. Confirmation of this
picture could come from
measurements at various beam energies,  allowing to vary the
local energy density
without  changing the collision  geometry.
Unfortunately, such a study  was not possible
at CERN.

What will happen at RHIC? The various models, when  extrapolated to
high energies, lead to a
severe  suppression, making even the observation of $J/\Psi$'s
problematic.  But these
extrapolations miss a potentially important feature. There is indeed
the interesting
possibility that
$J/\Psi$ production could be enhanced at RHIC, rather than
suppressed! The main idea
leading to this expectation is that, at the high energy of RHIC, several
$c\bar c$ pairs may be present in the same system. These pairs have a
finite probability
to recombine leading to a $J/\Psi$ yield growing as the square of
the number of $c\bar c$
pairs produced, and various calculations suggest that
this may  overcome the expected severe
suppression \cite{Thews:2001rj,Braun-Munzinger:2000px}.  Clearly it
will be very interesting to see the first data on $J/\Psi$
production  at RHIC!

\subsection{Jet quenching}

The last specific signal that I wish to discuss is a new one in the
list. It concerns indeed
an effect which was not observed at the SPS and which is related to the
energy loss of fast partons moving through a quark gluon plasma. The
calculation of this
effect  has been done by  various teams
\cite{Wang:1991ht,Zakharov:1996fv,Baier:1997kr,Wiedemann:2000ez,Gyulassy:2000zd},
using various formalisms,  leading to successive improvements in the
estimate, in particular
of its energy dependence (see
the talks by Levai,  Wang and Wiedemann).

The energy loss of a fast moving quark or gluon in a quark-gluon 
plasma is mostly due to
induced  gluon radiation and it
affects the  momentum distributions of the produced hadrons: when the
jets hadronize they
have less energy than they would if they had been produced in free
space. As a result, the
hadron momentum distribution  is softer.
Much excitement during the meeting was generated in particular by the 
first reports
on the $\pi_0$ spectra  by PHENIX, because these spectra
are  distorted in a way which
could be compatible with partonic energy loss. As Levai showed in his
contribution, while
perturbative QCD combined with a simple Glauber picture accounts
perfectly for the yields in
peripheral collisions, it  fails by a factor 10 for the most
central ones.
Jet quenching was proposed as a possible explanation and indeed a fit 
to the data can be
achieved  with a  moderate value of the energy loss,
$dE/dx\approx 0.25
$Gev/fm.

Is this a hint for the presence of a quark-gluon plasma? Data
with higher transverse momentum particles will certainly bring 
further crucial information.
One should note here that the extrapolation from peripheral to 
central collisions
involve assumptions about the dependence of centrality on the number 
of binary collisions
which are perhaps not fully under control.  It should also be 
emphasized that this is an
indirect observation since we do not see the jets  and, as was the 
case for the phenomenon
of
$J/\Psi$ suppression, one will perhaps find hadronic mechanisms to 
explain the data. It
remains that  we have here a new and exciting phenomenon which, when 
fully understood,
will   provide   a new tool for exploring the properties of the 
produced matter. It will
complement nicely the use of other penetrating probes such as the $J/\Psi$ meson.

\section{CONCLUDING REMARKS}

At the end of such a meeting, one cannot help feeling grateful to all 
the people
  who have contributed to make
RHIC available for the study of dense and hot matter. What has been a dream
for many years starts to be realized, perhaps beyond expectations, and the
new  perspectives that
RHIC opens in the field are quite exciting.

Although many of the data presented during the week may just seem to
confirm  the overall picture that
has emerged from previous studies at lower energies, a few
observations may force us already to review certain aspects of our 
standard picture of high
energy nuclear collisions. There is  
evidence that, as
expected, quark and gluon degrees of freedom play a more visible role at
RHIC than at the SPS, and indeed QCD is being more directly involved in the
interpretation of the data.  Besides, a few new non trivial effects have
been reported, and new probes have  become available.   At the same time 
several exciting
theoretical developments suggest that the gap is closing 
between fundamental studies of QCD and 
   more phenomenological ones on the physics of heavy ions. In summary,
both the data which have been presented  and the progress in theory that
have been reported   will certainly  be the source of
inspiration for a lot of work in the months to come. This is the sign
of a very successful meeting!

\vspace{0.5cm}

\noindent{\bf Acknowledgments}. I wish to thank Jean-Yves Ollitrault
for numerous discussions
and for his critical reading of the manuscript.


\begin{thebibliography}{99}


\bibitem{Karsch:2000kv}
F.~Karsch, E.~Laermann and A.~Peikert,
hep-lat/0012023.

\bibitem{Peshier:1996ty}
A.~Peshier, B.~Kampfer, O.~P.~Pavlenko and G.~Soff,
Phys.\ Rev.\ D {\bf 54} (1996) 2399.

\bibitem{Blaizot:1999ip}
J.~P.~Blaizot, E.~Iancu and A.~Rebhan,
Phys.\ Rev.\ Lett.\ {\bf 83} (1999) 2906
[hep-ph/9906340];
Phys.\ Lett.\ B {\bf 470} (1999) 181
[hep-ph/9910309];
Phys.\ Rev.\ D {\bf 63} (2001) 065003
[hep-ph/0005003].

\bibitem{Andersen:1999fw}
J.~O.~Andersen, E.~Braaten and M.~Strickland,
Phys.\ Rev.\ Lett.\ {\bf 83} (1999) 2139
[hep-ph/9902327];
Phys.\ Rev.\ D {\bf 61} (2000) 014017
[hep-ph/9905337];
Phys.\ Rev.\ D {\bf 61} (2000) 074016
[hep-ph/9908323].

\bibitem{Kajantie:2001iz}
K.~Kajantie, M.~Laine, K.~Rummukainen and Y.~Schroder,
Phys.\ Rev.\ Lett.\ {\bf 86} (2001) 10
[hep-ph/0007109].

\bibitem{Svetitsky:1982gs}
B.~Svetitsky and L.~G.~Yaffe,
Nucl.\ Phys.\ B {\bf 210} (1982) 423.


\bibitem{Pisarski:2000eq}
R.~D.~Pisarski,
Phys.\ Rev.\ D {\bf 62} (2000) 111501
[hep-ph/0006205].


\bibitem{Asakawa:2000tr}
M.~Asakawa, T.~Hatsuda and Y.~Nakahara,
hep-lat/0011040.

\bibitem{Alford:2001ze}
M.~Alford, J.~Bowers and K.~Rajagopal,
Phys.\ Rev.\ D {\bf 63} (2001) 074016
[hep-ph/0008208].

\bibitem{Cleymans99}
J.~Cleymans and K.~Redlich, nucl-th/9903063.

\bibitem{Becattini:1996if}
F.~Becattini,
Z.\ Phys.\ C {\bf 69} (1996) 485.

\bibitem{Gazdzicki:1999rk}
M.~Gazdzicki and M.~I.~Gorenstein,
Phys.\ Rev.\ Lett.\  {\bf 83} (1999) 4009
[hep-ph/9905515].

\bibitem{Eskola:2000my}
K.~J.~Eskola and K.~Tuominen,
hep-ph/0010319.

\bibitem{Bialas:1976ed}
A.~Bialas, M.~Bleszynski and W.~Czyz,
Nucl.\ Phys.\ B {\bf 111} (1976) 461.

\bibitem{Capella:1994yb}
A.~Capella, U.~Sukhatme, C.I.~Tan and J.~Tran Thanh Van,
Phys.\ Rept.\ {\bf 236} (1994) 225.

\bibitem{Kharzeev:2000ph}
D.~Kharzeev and M.~Nardi,
nucl-th/0012025.

\bibitem{Mueller:1999wm}
A.~H.~Mueller,
Nucl.\ Phys.\ B {\bf 558} (1999) 285
[hep-ph/9904404].

\bibitem{Blaizot:2001nr}
J.~Blaizot and E.~Iancu,
hep-ph/0101103.

\bibitem{Blaizot:1987nc}
J.~P.~Blaizot and A.~H.~Mueller,
Nucl.\ Phys.\ B {\bf 289} (1987) 847.

\bibitem{Mueller:1986wy}
A.~H.~Mueller and J.~Qiu,
Nucl.\ Phys.\ B {\bf 268} (1986) 427.

\bibitem{MCLRV94}
L.~McLerran and R.~Venugopalan, \PRD{49}{1994}{2233}; \PRD{49}{1994}{3352};
\PRD{50}{1994}{2225}.

\bibitem{AHMueller99}
A.H.~Mueller, ``Toward Equilibration in the Early Stages After a High
Energy Heavy
Ion Collision'', hep-ph/9906322.


\bibitem{Baier:2001sb}
R.~Baier, A.~H.~Mueller, D.~Schiff and D.~T.~Son,
Phys.\ Lett.\ B {\bf 502} (2001) 51
[hep-ph/0009237].

\bibitem{Asakawa00}
M.~Asakawa, U. Heinz, B. M\"uller, \PRL{85}{2000}{2072}; S. Jeon and V. Koch,
\PRL{85}{2000}{2076}.


\bibitem{Ollitrault:1992bk}
J.-Y.~Ollitrault,
Phys.\ Rev.\ D {\bf 46} (1992) 229;
Nucl.\ Phys.\ A {\bf 638} (1998) 195C
[nucl-ex/9802005].

\bibitem{Kolb:2001fh}
P.~F.~Kolb, P.~Huovinen, U.~Heinz and H.~Heiselberg,
Phys.\ Lett.\ B {\bf 500} (2001) 232
[hep-ph/0012137].

\bibitem{Huovinen:2001cy}
P.~Huovinen, P.~F.~Kolb, U.~Heinz, P.~V.~Ruuskanen and S.~A.~Voloshin,
Phys.\ Lett.\ B {\bf 503} (2001) 58
[hep-ph/0101136].

\bibitem{Wang:2001fq}
X.~Wang,
Phys.\ Rev.\ C {\bf 63} (2001) 054902
[nucl-th/0009019];
M.~Gyulassy, I.~Vitev and X.~N.~Wang,
Phys.\ Rev.\ Lett.\ {\bf 86} (2001) 2537
[nucl-th/0012092].

\bibitem{Brown00}
G.E.~Brown, M.~Rho and C.~Song, nucl-th/0010008.


\bibitem{Hamieh:2000tk}
S.~Hamieh, K.~Redlich and A.~Tounsi,
Phys.\ Lett.\ B {\bf 486} (2000) 61
[hep-ph/0006024].

\bibitem{Rapp:2000zw}
R.~Rapp and E.~Shuryak,
Phys.\ Lett.\ B {\bf 473} (2000) 13
[hep-ph/9909348].

\bibitem{Blaizot:2000ev}
J.~Blaizot, M.~Dinh and J.~Ollitrault,
Phys.\ Rev.\ Lett.\  {\bf 85} (2000) 4012
[nucl-th/0007020].

\bibitem{Thews:2001rj}
R.~L.~Thews, M.~Schroedter and J.~Rafelski,
Phys.\ Rev.\ C {\bf 63} (2001) 054905
[hep-ph/0007323].


\bibitem{Braun-Munzinger:2000px}
P.~Braun-Munzinger and J.~Stachel,
Phys.\ Lett.\ B {\bf 490} (2000) 196
[nucl-th/0007059].



\bibitem{Wang:1991ht}
X.~Wang and M.~Gyulassy,
Phys.\ Rev.\ D {\bf 44} (1991) 3501.

\bibitem{Zakharov:1996fv}
B.~G.~Zakharov,
JETP Lett.\ {\bf 63} (1996) 952
[hep-ph/9607440].

\bibitem{Baier:1997kr}
R.~Baier, Y.~L.~Dokshitzer, A.~H.~Mueller, S.~Peigne and D.~Schiff,
Nucl.\ Phys.\ B {\bf 483} (1997) 291
[hep-ph/9607355].

\bibitem{Wiedemann:2000ez}
U.~A.~Wiedemann,
Nucl.\ Phys.\ B {\bf 582} (2000) 409
[hep-ph/0003021].

\bibitem{Gyulassy:2000zd}
M.~Gyulassy, P.~Levai and I.~Vitev,
Nucl.\ Phys.\ B {\bf 571} (2000) 197
[hep-ph/9907461].


\end{thebibliography}
\end{document}